\documentclass[aps,twocolumn]{revtex4}
\usepackage{bm}
\begin{document}
\input{psfig}

\title{Comment on the use of the method of images for calculating
  electromagnetic responses of interacting spheres}
 
\author{Vadim A. Markel}

\affiliation{Departments of Radiology and Bioengineering, University
  of Pennsylvania, Philadelphia, PA 19104} 

\begin{abstract}
  In this Comment I argue that the method of images used by Huang, Yu,
  Gu and co-authors [Phys Rev. E, 65, 21401 (2002); Phys Rev. E., 69,
  51402 (2004)] to calculate electromagnetic properties of interacting
  spheres at finite frequencies is inapplicable and does not provide a
  physically meaningful approximation.
\end{abstract}

\date{\today} 
\maketitle

Recently, Huang, Yu, Gu and co-authors (referred to as HYG below) have
applied the method of images to study theoretically the
electromagnetic properties of two interacting spherical
particles~\cite{huang_02_1,huang_04_1}. As is well known, the method
of images can be applied to spherical conductors in the electrostatic
limit, i.e., when the dielectric constant $\epsilon$ can be formally
set to $i\infty$ and the Bergman-Milton spectral parameter
$s=1/(\epsilon-1)$ is equal to zero.  At finite frequencies, when $s$
is not small compared to the generalized depolarization factors $s_n$,
the method of images is not applicable.  However, HYG apply the method
to dielectric particles at arbitrary frequencies, assuming only that
the size of the two-sphere dimer is much smaller than the external
wavelength.  In particular, they claim to be able to extract the
factors $s_n$ and the corresponding oscillator strengths $F_n$, which
characterize the electromagnetic response of a system
within the quasistatics. In the first paper of the
series~\cite{yu_00_1} and in Ref.~\cite{huang_02_1} the authors
mention that their method is approximate. However, in the more recent
paper~\cite{huang_04_1} it is presented as exact and used without
restriction. In the present Comment I show that it is impossible to
calculate the quantities $s_n$ and $F_n$ using the method of images.
Moreover, the expressions for $s_n$ and $F_n$ given by HYG are not
consistent with the exact electrostatic solution.  Thus, the
mathematical formalism developed by HYG is not only not exact, but
does not provide a physically meaningful and controllable
approximation.

We start with a brief review of mathematical formalism used by HYG.
Within the quasistatics, dipole moment of an arbitrary particle
characterized by the dielectric function $\epsilon(\omega)$ and
excited by a homogeneous external field ${\bm E}_0\exp(-i\omega t)$
can be written as ${\bm d}\exp(-i\omega t)$ where ${\bm d} =
\hat{\alpha}{\bm E}_0$. Here $\hat{\alpha}$ is the polarizability
tensor. If polarization of the external field coincides with one of
the principal axes of $\hat{\alpha}$, both vectors ${\bm d}$ and ${\bm
  E}_0$ become collinear. The corresponding scalar polarizability can
be written in the Bergman-Milton spectral
representation~\cite{bergman_pr} as

\begin{equation}
\alpha = {v \over {4\pi}} \sum_n {{F_n} \over {s + s_n}} \ ,
\label{alpha_spectral}
\end{equation}

\noindent
where $v$ is the volume of the particle, $s_n$ - the generalized
depolarization factors satisfying $0<s_n<1$ and $F_n$ are the
corresponding oscillator strengths. 

In the case of two spheres, one principal axis of the polarizability
tensor coincides with the axis of symmetry, and the other two axes are
perpendicular to the first one and to each other, but otherwise
arbitrary. HYG consider two interacting spheres of the radius $R$ each
separated by the center-to-center distance $2L$, obtain the diagonal
elements of the polarizability tensor and derive the following
expressions for $F_n$ and $s_n$:

\begin{eqnarray}
\label{F_n}
&& F_n^{(L)}= F_n^{(T)}=F_n = \nonumber \\
&& \hspace{2cm} 4n(n+1)\sinh^3a \exp[-(2n+1)a] \ ,
\\
\label{s_n_L}
&& s_n^{(L)} = {1 \over 3}\left\{1 - 2\exp[-(2n+1)a] \right\} \ ,
\\
\label{s_n_T}
&& s_n^{(T)} = {1 \over 3}\left\{1 + \exp[-(2n+1)a] \right\} \ , \\
&& \hspace{4cm} n=1,2,3,\ldots \nonumber
\end{eqnarray}

\noindent
where the upper index $(L)$ denotes longitudinal modes, $(T)$ denotes
transverse modes, and $a$ is the solution to $\cosh a=L/R$, or,
explicitly, $a= \ln[L/R+\sqrt{(L/R)^2-1}]$~\cite{fn2}. It can be
verified that $F_n$ satisfy the sum rule $\sum_n F_n = 1$.

Everywhere below we consider only the longitudinal modes, although the
results of HYG for the transverse modes are also incorrect. The
longitudinal modes are more important physically, since they are known
to produce extremely high field enhancements in axially-symmetrical
arrays of nanospheres~\cite{li_03_1} and have been extensively studied
in conjunction with the single-molecule
spectroscopy~\cite{jiang_j_03_1}.

First, let us discuss the small-frequency limit for conductors. In
this limit, the dielectric function can be written as $\epsilon=4\pi
i\sigma/\omega$, where $\sigma$ is the static conductivity.
Correspondingly, $s \propto \omega \rightarrow 0$, and we can expand
$\alpha$ into a power series in $s$. The expansion can be obtained
from (\ref{alpha_spectral}) and reads

\begin{eqnarray}
\label{alpha_small_s}
&& \alpha = {v \over {4\pi}} \sum_{k=0}^\infty A_k s^k \ , \\
\label{Ck_def}
&& A_k = \sum_n F_n/s_n^{k+1} \ .
\end{eqnarray}

\noindent
The electrostatic polarizability is given by $\alpha_{\rm
  es}=(v/4\pi)A_0$. The method of images can provide an exact
expression for $\alpha_{\rm es}$ and, correspondingly, for $A_0$.
However, since there is an infinite number of different sets of
${F_n,s_n}$ that produce the same value of $A_0$, it is impossible to
find these coefficients from the electrostatic solution. We emphasize
that this is not possible even if one considers the inter-sphere
separation as an additional degree of freedom, since all quantities
($A_k,F_n$ and $s_n$) depend parametrically on $L/R$.  Instead, if the
summation in the right-hand side of (\ref{Ck_def}) is truncated at
$n=N$, one needs to calculate all coefficients $A_k$ from $k=0$ to
$k=2N-1$ in order to make the system of equations (\ref{Ck_def})
sufficiently determined. But the electrostatic solution based on the
method of images can provide only one of these coefficients, namely,
$A_0$.

\centerline{
\begingroup%
  \makeatletter%
  \newcommand{\GNUPLOTspecial}{%
    \@sanitize\catcode`\%=14\relax\special}%
  \setlength{\unitlength}{0.1bp}%
{\GNUPLOTspecial{!
/gnudict 256 dict def
gnudict begin
/Color false def
/Solid false def
/gnulinewidth 5.000 def
/userlinewidth gnulinewidth def
/vshift -33 def
/dl {10 mul} def
/hpt_ 31.5 def
/vpt_ 31.5 def
/hpt hpt_ def
/vpt vpt_ def
/M {moveto} bind def
/L {lineto} bind def
/R {rmoveto} bind def
/V {rlineto} bind def
/vpt2 vpt 2 mul def
/hpt2 hpt 2 mul def
/Lshow { currentpoint stroke M
  0 vshift R show } def
/Rshow { currentpoint stroke M
  dup stringwidth pop neg vshift R show } def
/Cshow { currentpoint stroke M
  dup stringwidth pop -2 div vshift R show } def
/UP { dup vpt_ mul /vpt exch def hpt_ mul /hpt exch def
  /hpt2 hpt 2 mul def /vpt2 vpt 2 mul def } def
/DL { Color {setrgbcolor Solid {pop []} if 0 setdash }
 {pop pop pop Solid {pop []} if 0 setdash} ifelse } def
/BL { stroke userlinewidth 2 mul setlinewidth } def
/AL { stroke userlinewidth 2 div setlinewidth } def
/UL { dup gnulinewidth mul /userlinewidth exch def
      dup 1 lt {pop 1} if 10 mul /udl exch def } def
/PL { stroke userlinewidth setlinewidth } def
/LTb { BL [] 0 0 0 DL } def
/LTa { AL [1 udl mul 2 udl mul] 0 setdash 0 0 0 setrgbcolor } def
/LT0 { PL [] 1 0 0 DL } def
/LT1 { PL [4 dl 2 dl] 0 1 0 DL } def
/LT2 { PL [2 dl 3 dl] 0 0 1 DL } def
/LT3 { PL [1 dl 1.5 dl] 1 0 1 DL } def
/LT4 { PL [5 dl 2 dl 1 dl 2 dl] 0 1 1 DL } def
/LT5 { PL [4 dl 3 dl 1 dl 3 dl] 1 1 0 DL } def
/LT6 { PL [2 dl 2 dl 2 dl 4 dl] 0 0 0 DL } def
/LT7 { PL [2 dl 2 dl 2 dl 2 dl 2 dl 4 dl] 1 0.3 0 DL } def
/LT8 { PL [2 dl 2 dl 2 dl 2 dl 2 dl 2 dl 2 dl 4 dl] 0.5 0.5 0.5 DL } def
/Pnt { stroke [] 0 setdash
   gsave 1 setlinecap M 0 0 V stroke grestore } def
/Dia { stroke [] 0 setdash 2 copy vpt add M
  hpt neg vpt neg V hpt vpt neg V
  hpt vpt V hpt neg vpt V closepath stroke
  Pnt } def
/Pls { stroke [] 0 setdash vpt sub M 0 vpt2 V
  currentpoint stroke M
  hpt neg vpt neg R hpt2 0 V stroke
  } def
/Box { stroke [] 0 setdash 2 copy exch hpt sub exch vpt add M
  0 vpt2 neg V hpt2 0 V 0 vpt2 V
  hpt2 neg 0 V closepath stroke
  Pnt } def
/Crs { stroke [] 0 setdash exch hpt sub exch vpt add M
  hpt2 vpt2 neg V currentpoint stroke M
  hpt2 neg 0 R hpt2 vpt2 V stroke } def
/TriU { stroke [] 0 setdash 2 copy vpt 1.12 mul add M
  hpt neg vpt -1.62 mul V
  hpt 2 mul 0 V
  hpt neg vpt 1.62 mul V closepath stroke
  Pnt  } def
/Star { 2 copy Pls Crs } def
/BoxF { stroke [] 0 setdash exch hpt sub exch vpt add M
  0 vpt2 neg V  hpt2 0 V  0 vpt2 V
  hpt2 neg 0 V  closepath fill } def
/TriUF { stroke [] 0 setdash vpt 1.12 mul add M
  hpt neg vpt -1.62 mul V
  hpt 2 mul 0 V
  hpt neg vpt 1.62 mul V closepath fill } def
/TriD { stroke [] 0 setdash 2 copy vpt 1.12 mul sub M
  hpt neg vpt 1.62 mul V
  hpt 2 mul 0 V
  hpt neg vpt -1.62 mul V closepath stroke
  Pnt  } def
/TriDF { stroke [] 0 setdash vpt 1.12 mul sub M
  hpt neg vpt 1.62 mul V
  hpt 2 mul 0 V
  hpt neg vpt -1.62 mul V closepath fill} def
/DiaF { stroke [] 0 setdash vpt add M
  hpt neg vpt neg V hpt vpt neg V
  hpt vpt V hpt neg vpt V closepath fill } def
/Pent { stroke [] 0 setdash 2 copy gsave
  translate 0 hpt M 4 {72 rotate 0 hpt L} repeat
  closepath stroke grestore Pnt } def
/PentF { stroke [] 0 setdash gsave
  translate 0 hpt M 4 {72 rotate 0 hpt L} repeat
  closepath fill grestore } def
/Circle { stroke [] 0 setdash 2 copy
  hpt 0 360 arc stroke Pnt } def
/CircleF { stroke [] 0 setdash hpt 0 360 arc fill } def
/C0 { BL [] 0 setdash 2 copy moveto vpt 90 450  arc } bind def
/C1 { BL [] 0 setdash 2 copy        moveto
       2 copy  vpt 0 90 arc closepath fill
               vpt 0 360 arc closepath } bind def
/C2 { BL [] 0 setdash 2 copy moveto
       2 copy  vpt 90 180 arc closepath fill
               vpt 0 360 arc closepath } bind def
/C3 { BL [] 0 setdash 2 copy moveto
       2 copy  vpt 0 180 arc closepath fill
               vpt 0 360 arc closepath } bind def
/C4 { BL [] 0 setdash 2 copy moveto
       2 copy  vpt 180 270 arc closepath fill
               vpt 0 360 arc closepath } bind def
/C5 { BL [] 0 setdash 2 copy moveto
       2 copy  vpt 0 90 arc
       2 copy moveto
       2 copy  vpt 180 270 arc closepath fill
               vpt 0 360 arc } bind def
/C6 { BL [] 0 setdash 2 copy moveto
      2 copy  vpt 90 270 arc closepath fill
              vpt 0 360 arc closepath } bind def
/C7 { BL [] 0 setdash 2 copy moveto
      2 copy  vpt 0 270 arc closepath fill
              vpt 0 360 arc closepath } bind def
/C8 { BL [] 0 setdash 2 copy moveto
      2 copy vpt 270 360 arc closepath fill
              vpt 0 360 arc closepath } bind def
/C9 { BL [] 0 setdash 2 copy moveto
      2 copy  vpt 270 450 arc closepath fill
              vpt 0 360 arc closepath } bind def
/C10 { BL [] 0 setdash 2 copy 2 copy moveto vpt 270 360 arc closepath fill
       2 copy moveto
       2 copy vpt 90 180 arc closepath fill
               vpt 0 360 arc closepath } bind def
/C11 { BL [] 0 setdash 2 copy moveto
       2 copy  vpt 0 180 arc closepath fill
       2 copy moveto
       2 copy  vpt 270 360 arc closepath fill
               vpt 0 360 arc closepath } bind def
/C12 { BL [] 0 setdash 2 copy moveto
       2 copy  vpt 180 360 arc closepath fill
               vpt 0 360 arc closepath } bind def
/C13 { BL [] 0 setdash  2 copy moveto
       2 copy  vpt 0 90 arc closepath fill
       2 copy moveto
       2 copy  vpt 180 360 arc closepath fill
               vpt 0 360 arc closepath } bind def
/C14 { BL [] 0 setdash 2 copy moveto
       2 copy  vpt 90 360 arc closepath fill
               vpt 0 360 arc } bind def
/C15 { BL [] 0 setdash 2 copy vpt 0 360 arc closepath fill
               vpt 0 360 arc closepath } bind def
/Rec   { newpath 4 2 roll moveto 1 index 0 rlineto 0 exch rlineto
       neg 0 rlineto closepath } bind def
/Square { dup Rec } bind def
/Bsquare { vpt sub exch vpt sub exch vpt2 Square } bind def
/S0 { BL [] 0 setdash 2 copy moveto 0 vpt rlineto BL Bsquare } bind def
/S1 { BL [] 0 setdash 2 copy vpt Square fill Bsquare } bind def
/S2 { BL [] 0 setdash 2 copy exch vpt sub exch vpt Square fill Bsquare } bind def
/S3 { BL [] 0 setdash 2 copy exch vpt sub exch vpt2 vpt Rec fill Bsquare } bind def
/S4 { BL [] 0 setdash 2 copy exch vpt sub exch vpt sub vpt Square fill Bsquare } bind def
/S5 { BL [] 0 setdash 2 copy 2 copy vpt Square fill
       exch vpt sub exch vpt sub vpt Square fill Bsquare } bind def
/S6 { BL [] 0 setdash 2 copy exch vpt sub exch vpt sub vpt vpt2 Rec fill Bsquare } bind def
/S7 { BL [] 0 setdash 2 copy exch vpt sub exch vpt sub vpt vpt2 Rec fill
       2 copy vpt Square fill
       Bsquare } bind def
/S8 { BL [] 0 setdash 2 copy vpt sub vpt Square fill Bsquare } bind def
/S9 { BL [] 0 setdash 2 copy vpt sub vpt vpt2 Rec fill Bsquare } bind def
/S10 { BL [] 0 setdash 2 copy vpt sub vpt Square fill 2 copy exch vpt sub exch vpt Square fill
       Bsquare } bind def
/S11 { BL [] 0 setdash 2 copy vpt sub vpt Square fill 2 copy exch vpt sub exch vpt2 vpt Rec fill
       Bsquare } bind def
/S12 { BL [] 0 setdash 2 copy exch vpt sub exch vpt sub vpt2 vpt Rec fill Bsquare } bind def
/S13 { BL [] 0 setdash 2 copy exch vpt sub exch vpt sub vpt2 vpt Rec fill
       2 copy vpt Square fill Bsquare } bind def
/S14 { BL [] 0 setdash 2 copy exch vpt sub exch vpt sub vpt2 vpt Rec fill
       2 copy exch vpt sub exch vpt Square fill Bsquare } bind def
/S15 { BL [] 0 setdash 2 copy Bsquare fill Bsquare } bind def
/D0 { gsave translate 45 rotate 0 0 S0 stroke grestore } bind def
/D1 { gsave translate 45 rotate 0 0 S1 stroke grestore } bind def
/D2 { gsave translate 45 rotate 0 0 S2 stroke grestore } bind def
/D3 { gsave translate 45 rotate 0 0 S3 stroke grestore } bind def
/D4 { gsave translate 45 rotate 0 0 S4 stroke grestore } bind def
/D5 { gsave translate 45 rotate 0 0 S5 stroke grestore } bind def
/D6 { gsave translate 45 rotate 0 0 S6 stroke grestore } bind def
/D7 { gsave translate 45 rotate 0 0 S7 stroke grestore } bind def
/D8 { gsave translate 45 rotate 0 0 S8 stroke grestore } bind def
/D9 { gsave translate 45 rotate 0 0 S9 stroke grestore } bind def
/D10 { gsave translate 45 rotate 0 0 S10 stroke grestore } bind def
/D11 { gsave translate 45 rotate 0 0 S11 stroke grestore } bind def
/D12 { gsave translate 45 rotate 0 0 S12 stroke grestore } bind def
/D13 { gsave translate 45 rotate 0 0 S13 stroke grestore } bind def
/D14 { gsave translate 45 rotate 0 0 S14 stroke grestore } bind def
/D15 { gsave translate 45 rotate 0 0 S15 stroke grestore } bind def
/DiaE { stroke [] 0 setdash vpt add M
  hpt neg vpt neg V hpt vpt neg V
  hpt vpt V hpt neg vpt V closepath stroke } def
/BoxE { stroke [] 0 setdash exch hpt sub exch vpt add M
  0 vpt2 neg V hpt2 0 V 0 vpt2 V
  hpt2 neg 0 V closepath stroke } def
/TriUE { stroke [] 0 setdash vpt 1.12 mul add M
  hpt neg vpt -1.62 mul V
  hpt 2 mul 0 V
  hpt neg vpt 1.62 mul V closepath stroke } def
/TriDE { stroke [] 0 setdash vpt 1.12 mul sub M
  hpt neg vpt 1.62 mul V
  hpt 2 mul 0 V
  hpt neg vpt -1.62 mul V closepath stroke } def
/PentE { stroke [] 0 setdash gsave
  translate 0 hpt M 4 {72 rotate 0 hpt L} repeat
  closepath stroke grestore } def
/CircE { stroke [] 0 setdash 
  hpt 0 360 arc stroke } def
/Opaque { gsave closepath 1 setgray fill grestore 0 setgray closepath } def
/DiaW { stroke [] 0 setdash vpt add M
  hpt neg vpt neg V hpt vpt neg V
  hpt vpt V hpt neg vpt V Opaque stroke } def
/BoxW { stroke [] 0 setdash exch hpt sub exch vpt add M
  0 vpt2 neg V hpt2 0 V 0 vpt2 V
  hpt2 neg 0 V Opaque stroke } def
/TriUW { stroke [] 0 setdash vpt 1.12 mul add M
  hpt neg vpt -1.62 mul V
  hpt 2 mul 0 V
  hpt neg vpt 1.62 mul V Opaque stroke } def
/TriDW { stroke [] 0 setdash vpt 1.12 mul sub M
  hpt neg vpt 1.62 mul V
  hpt 2 mul 0 V
  hpt neg vpt -1.62 mul V Opaque stroke } def
/PentW { stroke [] 0 setdash gsave
  translate 0 hpt M 4 {72 rotate 0 hpt L} repeat
  Opaque stroke grestore } def
/CircW { stroke [] 0 setdash 
  hpt 0 360 arc Opaque stroke } def
/BoxFill { gsave Rec 1 setgray fill grestore } def
/Symbol-Oblique /Symbol findfont [1 0 .167 1 0 0] makefont
dup length dict begin {1 index /FID eq {pop pop} {def} ifelse} forall
currentdict end definefont pop
end
}}%
\begin{picture}(2520,1512)(0,0)%
{\GNUPLOTspecial{"
gnudict begin
gsave
0 0 translate
0.100 0.100 scale
0 setgray
newpath
1.000 UL
LTb
400 200 M
63 0 V
1907 0 R
-63 0 V
400 322 M
31 0 V
1939 0 R
-31 0 V
400 393 M
31 0 V
1939 0 R
-31 0 V
400 443 M
31 0 V
1939 0 R
-31 0 V
400 482 M
31 0 V
1939 0 R
-31 0 V
400 514 M
31 0 V
1939 0 R
-31 0 V
400 541 M
31 0 V
1939 0 R
-31 0 V
400 565 M
31 0 V
1939 0 R
-31 0 V
400 586 M
31 0 V
1939 0 R
-31 0 V
400 604 M
63 0 V
1907 0 R
-63 0 V
400 726 M
31 0 V
1939 0 R
-31 0 V
400 797 M
31 0 V
1939 0 R
-31 0 V
400 847 M
31 0 V
1939 0 R
-31 0 V
400 886 M
31 0 V
1939 0 R
-31 0 V
400 918 M
31 0 V
1939 0 R
-31 0 V
400 945 M
31 0 V
1939 0 R
-31 0 V
400 969 M
31 0 V
1939 0 R
-31 0 V
400 990 M
31 0 V
1939 0 R
-31 0 V
400 1008 M
63 0 V
1907 0 R
-63 0 V
400 1130 M
31 0 V
1939 0 R
-31 0 V
400 1201 M
31 0 V
1939 0 R
-31 0 V
400 1251 M
31 0 V
1939 0 R
-31 0 V
400 1290 M
31 0 V
1939 0 R
-31 0 V
400 1322 M
31 0 V
1939 0 R
-31 0 V
400 1349 M
31 0 V
1939 0 R
-31 0 V
400 1373 M
31 0 V
1939 0 R
-31 0 V
400 1394 M
31 0 V
1939 0 R
-31 0 V
400 1412 M
63 0 V
1907 0 R
-63 0 V
400 200 M
0 63 V
0 1149 R
0 -63 V
598 200 M
0 31 V
0 1181 R
0 -31 V
713 200 M
0 31 V
0 1181 R
0 -31 V
795 200 M
0 31 V
0 1181 R
0 -31 V
859 200 M
0 31 V
0 1181 R
0 -31 V
911 200 M
0 31 V
0 1181 R
0 -31 V
955 200 M
0 31 V
0 1181 R
0 -31 V
993 200 M
0 31 V
0 1181 R
0 -31 V
1027 200 M
0 31 V
0 1181 R
0 -31 V
1057 200 M
0 63 V
0 1149 R
0 -63 V
1254 200 M
0 31 V
0 1181 R
0 -31 V
1370 200 M
0 31 V
0 1181 R
0 -31 V
1452 200 M
0 31 V
0 1181 R
0 -31 V
1516 200 M
0 31 V
0 1181 R
0 -31 V
1568 200 M
0 31 V
0 1181 R
0 -31 V
1612 200 M
0 31 V
0 1181 R
0 -31 V
1650 200 M
0 31 V
0 1181 R
0 -31 V
1683 200 M
0 31 V
0 1181 R
0 -31 V
1713 200 M
0 63 V
0 1149 R
0 -63 V
1911 200 M
0 31 V
0 1181 R
0 -31 V
2027 200 M
0 31 V
0 1181 R
0 -31 V
2109 200 M
0 31 V
0 1181 R
0 -31 V
2172 200 M
0 31 V
0 1181 R
0 -31 V
2224 200 M
0 31 V
0 1181 R
0 -31 V
2268 200 M
0 31 V
0 1181 R
0 -31 V
2306 200 M
0 31 V
0 1181 R
0 -31 V
2340 200 M
0 31 V
0 1181 R
0 -31 V
2370 200 M
0 63 V
0 1149 R
0 -63 V
1.000 UL
LTb
400 200 M
1970 0 V
0 1212 V
-1970 0 V
400 200 L
1.000 UL
LT0
2007 1299 M
263 0 V
400 1107 M
20 -2 V
19 -3 V
20 -2 V
20 -2 V
19 -3 V
20 -3 V
20 -2 V
20 -3 V
19 -3 V
20 -3 V
20 -3 V
19 -3 V
20 -3 V
20 -4 V
19 -3 V
20 -3 V
20 -4 V
20 -4 V
19 -4 V
20 -3 V
20 -5 V
19 -4 V
20 -4 V
20 -4 V
19 -5 V
20 -5 V
20 -5 V
20 -5 V
19 -5 V
20 -5 V
20 -6 V
19 -6 V
20 -6 V
20 -6 V
19 -6 V
20 -7 V
20 -6 V
20 -7 V
19 -8 V
20 -7 V
20 -8 V
19 -8 V
20 -8 V
20 -9 V
19 -9 V
20 -9 V
20 -10 V
20 -10 V
19 -10 V
20 -11 V
20 -11 V
19 -11 V
20 -12 V
20 -12 V
20 -13 V
19 -13 V
20 -14 V
20 -14 V
19 -15 V
20 -15 V
20 -15 V
19 -17 V
20 -16 V
20 -17 V
19 -18 V
20 -18 V
20 -19 V
20 -20 V
19 -20 V
20 -20 V
20 -21 V
19 -22 V
20 -22 V
20 -23 V
19 -23 V
20 -24 V
20 -24 V
20 -25 V
19 -26 V
20 -26 V
20 -26 V
19 -27 V
20 -27 V
12 -17 V
1.000 UL
LT1
2007 1199 M
263 0 V
400 1384 M
1 -18 V
2 -15 V
1 -13 V
1 -13 V
2 -11 V
1 -10 V
1 -10 V
2 -8 V
1 -9 V
1 -7 V
1 -7 V
2 -7 V
1 -6 V
1 -6 V
2 -6 V
1 -6 V
1 -5 V
2 -5 V
1 -5 V
1 -4 V
2 -5 V
1 -4 V
1 -4 V
2 -4 V
1 -4 V
1 -3 V
1 -4 V
2 -3 V
1 -4 V
1 -3 V
2 -3 V
1 -3 V
1 -3 V
2 -3 V
1 -3 V
1 -3 V
2 -3 V
1 -3 V
1 -2 V
2 -3 V
1 -2 V
1 -3 V
1 -2 V
2 -3 V
1 -2 V
1 -3 V
2 -2 V
1 -2 V
1 -3 V
2 -2 V
1 -2 V
1 -2 V
2 -3 V
1 -2 V
1 -2 V
2 -2 V
1 -2 V
1 -2 V
1 -2 V
2 -2 V
1 -2 V
1 -3 V
2 -2 V
1 -2 V
1 -2 V
2 -2 V
1 -2 V
1 -2 V
2 -2 V
1 -2 V
1 -2 V
2 -2 V
1 -2 V
1 -2 V
2 -2 V
1 -2 V
1 -2 V
1 -2 V
2 -2 V
1 -2 V
1 -2 V
2 -2 V
1 -3 V
1 -2 V
2 -2 V
1 -2 V
1 -2 V
2 -2 V
1 -3 V
1 -2 V
2 -2 V
1 -2 V
1 -3 V
1 -2 V
2 -2 V
1 -3 V
1 -2 V
2 -3 V
1 -2 V
1 -3 V
2 -2 V
1 -3 V
1 -3 V
2 -2 V
1 -3 V
1 -3 V
2 -3 V
1 -3 V
1 -3 V
1 -3 V
2 -3 V
1 -3 V
1 -4 V
2 -3 V
1 -4 V
1 -3 V
2 -4 V
1 -4 V
1 -4 V
2 -4 V
1 -4 V
1 -4 V
2 -5 V
1 -4 V
1 -5 V
1 -5 V
2 -6 V
1 -5 V
1 -6 V
2 -6 V
1 -6 V
1 -7 V
2 -7 V
1 -8 V
1 -8 V
2 -8 V
1 -9 V
1 -10 V
2 -10 V
1 -12 V
1 -12 V
1 -14 V
2 -15 V
1 -17 V
1 -19 V
2 -21 V
1 -26 V
1 -31 V
2 -38 V
1 -50 V
1 -73 V
2 -132 V
96 1045 R
1 -12 V
1 -12 V
1 -11 V
2 -10 V
1 -10 V
1 -9 V
2 -8 V
1 -8 V
1 -7 V
2 -7 V
1 -7 V
1 -6 V
2 -6 V
1 -5 V
1 -6 V
2 -5 V
1 -5 V
1 -4 V
1 -5 V
2 -4 V
1 -4 V
1 -4 V
2 -4 V
1 -3 V
1 -4 V
2 -3 V
1 -4 V
1 -3 V
2 -3 V
1 -3 V
1 -3 V
2 -3 V
1 -3 V
1 -2 V
1 -3 V
2 -3 V
1 -2 V
1 -3 V
2 -2 V
1 -2 V
1 -2 V
2 -3 V
1 -2 V
1 -2 V
2 -2 V
1 -2 V
1 -2 V
2 -2 V
1 -2 V
1 -2 V
1 -2 V
2 -1 V
1 -2 V
1 -2 V
2 -2 V
1 -1 V
1 -2 V
2 -1 V
1 -2 V
1 -2 V
2 -1 V
1 -2 V
1 -1 V
2 -2 V
1 -1 V
1 -1 V
1 -2 V
2 -1 V
1 -2 V
1 -1 V
2 -1 V
1 -1 V
1 -2 V
2 -1 V
1 -1 V
1 -2 V
2 -1 V
1 -1 V
1 -1 V
2 -1 V
1 -1 V
1 -2 V
2 -1 V
1 -1 V
1 -1 V
1 -1 V
2 -1 V
1 -1 V
1 -1 V
2 -1 V
1 -1 V
1 -1 V
2 -1 V
1 -1 V
1 -1 V
2 -1 V
1 -1 V
1 -1 V
2 -1 V
1 -1 V
1 -1 V
1 -1 V
2 -1 V
1 -1 V
1 -1 V
2 -1 V
1 -1 V
1 -1 V
2 -1 V
1 0 V
1 -1 V
2 -1 V
1 -1 V
1 -1 V
2 -1 V
1 -1 V
1 0 V
1 -1 V
2 -1 V
1 -1 V
1 -1 V
2 -1 V
1 0 V
1 -1 V
2 -1 V
1 -1 V
1 -1 V
2 0 V
1 -1 V
1 -1 V
2 -1 V
1 0 V
1 -1 V
1 -1 V
2 -1 V
1 0 V
1 -1 V
2 -1 V
1 -1 V
1 0 V
2 -1 V
1 -1 V
1 0 V
2 -1 V
1 -1 V
1 -1 V
2 0 V
1 -1 V
1 -1 V
2 0 V
1 -1 V
1 -1 V
1 -1 V
2 0 V
1 -1 V
1 -1 V
2 0 V
1 -1 V
1 -1 V
2 0 V
1 -1 V
1 -1 V
2 0 V
1 -1 V
1 -1 V
2 0 V
1 -1 V
1 -1 V
1 0 V
2 -1 V
1 0 V
1 -1 V
2 -1 V
1 0 V
1 -1 V
2 -1 V
1 0 V
1 -1 V
2 -1 V
1 0 V
1 -1 V
2 0 V
1 -1 V
1 -1 V
1 0 V
2 -1 V
1 0 V
1 -1 V
2 -1 V
1 0 V
1 -1 V
2 -1 V
1 0 V
1 -1 V
2 0 V
1 -1 V
1 -1 V
2 0 V
1 -1 V
1 0 V
1 -1 V
2 -1 V
1 0 V
1 -1 V
2 0 V
1 -1 V
1 0 V
2 -1 V
1 -1 V
1 0 V
2 -1 V
1 0 V
1 -1 V
2 -1 V
1 0 V
1 -1 V
1 0 V
2 -1 V
1 0 V
1 -1 V
2 -1 V
1 0 V
1 -1 V
2 0 V
1 -1 V
1 -1 V
2 0 V
1 -1 V
1 0 V
2 -1 V
1 0 V
1 -1 V
2 -1 V
1 0 V
1 -1 V
1 0 V
2 -1 V
1 0 V
1 -1 V
2 0 V
1 -1 V
1 -1 V
2 0 V
1 -1 V
currentpoint stroke M
1 0 V
2 -1 V
1 0 V
1 -1 V
2 -1 V
1 0 V
1 -1 V
1 0 V
2 -1 V
1 0 V
1 -1 V
2 0 V
1 -1 V
1 -1 V
2 0 V
1 -1 V
1 0 V
2 -1 V
1 0 V
1 -1 V
2 0 V
1 -1 V
1 -1 V
1 0 V
2 -1 V
1 0 V
1 -1 V
2 0 V
1 -1 V
1 -1 V
2 0 V
1 -1 V
1 0 V
2 -1 V
1 0 V
1 -1 V
2 0 V
1 -1 V
1 -1 V
1 0 V
2 -1 V
1 0 V
1 -1 V
2 0 V
1 -1 V
1 0 V
2 -1 V
1 -1 V
1 0 V
2 -1 V
1 0 V
1 -1 V
2 0 V
1 -1 V
1 0 V
2 -1 V
1 -1 V
1 0 V
1 -1 V
2 0 V
1 -1 V
1 0 V
2 -1 V
1 0 V
1 -1 V
2 -1 V
1 0 V
1 -1 V
2 0 V
1 -1 V
1 0 V
2 -1 V
1 -1 V
1 0 V
1 -1 V
2 0 V
1 -1 V
1 0 V
2 -1 V
1 0 V
1 -1 V
2 -1 V
1 0 V
1 -1 V
2 0 V
1 -1 V
1 0 V
2 -1 V
1 -1 V
1 0 V
1 -1 V
2 0 V
1 -1 V
1 0 V
2 -1 V
1 -1 V
1 0 V
2 -1 V
1 0 V
1 -1 V
2 0 V
1 -1 V
1 -1 V
2 0 V
1 -1 V
1 0 V
1 -1 V
2 0 V
1 -1 V
1 -1 V
2 0 V
1 -1 V
1 0 V
2 -1 V
1 -1 V
1 0 V
2 -1 V
1 0 V
1 -1 V
2 0 V
1 -1 V
1 -1 V
1 0 V
2 -1 V
1 0 V
1 -1 V
2 -1 V
1 0 V
1 -1 V
2 0 V
1 -1 V
1 -1 V
2 0 V
1 -1 V
1 0 V
2 -1 V
1 -1 V
1 0 V
2 -1 V
1 0 V
1 -1 V
1 0 V
2 -1 V
1 -1 V
1 0 V
2 -1 V
1 -1 V
1 0 V
2 -1 V
1 0 V
1 -1 V
2 -1 V
1 0 V
1 -1 V
2 0 V
1 -1 V
1 -1 V
1 0 V
2 -1 V
1 0 V
1 -1 V
2 -1 V
1 0 V
1 -1 V
2 -1 V
1 0 V
1 -1 V
2 0 V
1 -1 V
1 -1 V
2 0 V
1 -1 V
1 -1 V
1 0 V
2 -1 V
1 0 V
1 -1 V
2 -1 V
1 0 V
1 -1 V
2 -1 V
1 0 V
1 -1 V
2 0 V
1 -1 V
1 -1 V
2 0 V
1 -1 V
1 -1 V
1 0 V
2 -1 V
1 -1 V
1 0 V
2 -1 V
1 -1 V
1 0 V
2 -1 V
1 0 V
1 -1 V
2 -1 V
1 0 V
1 -1 V
2 -1 V
1 0 V
1 -1 V
2 -1 V
1 0 V
1 -1 V
1 -1 V
2 0 V
1 -1 V
1 -1 V
2 0 V
1 -1 V
1 -1 V
2 0 V
1 -1 V
1 -1 V
2 0 V
1 -1 V
1 -1 V
2 0 V
1 -1 V
1 -1 V
1 0 V
2 -1 V
1 -1 V
1 -1 V
2 0 V
1 -1 V
1 -1 V
2 0 V
1 -1 V
1 -1 V
2 0 V
1 -1 V
1 -1 V
2 0 V
1 -1 V
1 -1 V
1 -1 V
2 0 V
1 -1 V
1 -1 V
2 0 V
1 -1 V
1 -1 V
2 -1 V
1 0 V
1 -1 V
2 -1 V
1 0 V
1 -1 V
2 -1 V
1 -1 V
1 0 V
1 -1 V
2 -1 V
1 0 V
1 -1 V
2 -1 V
1 -1 V
1 0 V
2 -1 V
1 -1 V
1 -1 V
2 0 V
1 -1 V
1 -1 V
2 0 V
1 -1 V
1 -1 V
1 -1 V
2 0 V
1 -1 V
1 -1 V
2 -1 V
1 0 V
1 -1 V
2 -1 V
1 -1 V
1 0 V
2 -1 V
1 -1 V
1 -1 V
2 -1 V
1 0 V
1 -1 V
2 -1 V
1 -1 V
1 0 V
1 -1 V
2 -1 V
1 -1 V
1 0 V
2 -1 V
1 -1 V
1 -1 V
2 -1 V
1 0 V
1 -1 V
2 -1 V
1 -1 V
1 -1 V
2 0 V
1 -1 V
1 -1 V
1 -1 V
2 -1 V
1 0 V
1 -1 V
2 -1 V
1 -1 V
1 -1 V
2 0 V
1 -1 V
1 -1 V
2 -1 V
1 -1 V
1 0 V
2 -1 V
1 -1 V
1 -1 V
1 -1 V
2 -1 V
1 0 V
1 -1 V
2 -1 V
1 -1 V
1 -1 V
2 -1 V
1 0 V
1 -1 V
2 -1 V
1 -1 V
1 -1 V
2 -1 V
1 0 V
1 -1 V
1 -1 V
2 -1 V
1 -1 V
1 -1 V
2 -1 V
1 0 V
1 -1 V
2 -1 V
1 -1 V
1 -1 V
2 -1 V
1 -1 V
1 -1 V
2 0 V
1 -1 V
1 -1 V
2 -1 V
1 -1 V
1 -1 V
1 -1 V
2 -1 V
1 0 V
1 -1 V
2 -1 V
1 -1 V
1 -1 V
2 -1 V
1 -1 V
1 -1 V
2 -1 V
1 -1 V
1 -1 V
2 0 V
1 -1 V
1 -1 V
1 -1 V
2 -1 V
1 -1 V
1 -1 V
2 -1 V
1 -1 V
1 -1 V
2 -1 V
1 -1 V
1 -1 V
2 0 V
1 -1 V
1 -1 V
2 -1 V
1 -1 V
1 -1 V
1 -1 V
2 -1 V
1 -1 V
1 -1 V
2 -1 V
1 -1 V
1 -1 V
2 -1 V
1 -1 V
1 -1 V
currentpoint stroke M
2 -1 V
1 -1 V
1 -1 V
2 -1 V
1 -1 V
1 -1 V
1 -1 V
2 -1 V
1 -1 V
1 -1 V
2 -1 V
1 -1 V
1 -1 V
2 -1 V
1 -1 V
1 -1 V
2 -1 V
1 -1 V
1 -1 V
2 -1 V
1 -1 V
1 -1 V
1 -1 V
2 -1 V
1 -1 V
1 -1 V
2 -1 V
1 -1 V
1 -1 V
2 -1 V
1 -1 V
1 -1 V
2 -1 V
1 -1 V
1 -1 V
2 -1 V
1 -1 V
1 -1 V
2 -1 V
1 -1 V
1 -1 V
1 -1 V
2 -1 V
1 -1 V
1 -1 V
2 -2 V
1 -1 V
1 -1 V
2 -1 V
1 -1 V
1 -1 V
2 -1 V
1 -1 V
1 -1 V
2 -1 V
1 -1 V
1 -1 V
1 -1 V
2 -2 V
1 -1 V
1 -1 V
2 -1 V
1 -1 V
1 -1 V
2 -1 V
1 -1 V
1 -1 V
2 -2 V
1 -1 V
1 -1 V
2 -1 V
1 -1 V
1 -1 V
1 -1 V
2 -1 V
1 -2 V
1 -1 V
2 -1 V
1 -1 V
1 -1 V
2 -1 V
1 -1 V
1 -2 V
2 -1 V
1 -1 V
1 -1 V
2 -1 V
1 -1 V
1 -1 V
1 -2 V
2 -1 V
1 -1 V
1 -1 V
2 -1 V
1 -2 V
1 -1 V
2 -1 V
1 -1 V
1 -1 V
2 -1 V
1 -2 V
1 -1 V
2 -1 V
1 -1 V
1 -1 V
2 -2 V
1 -1 V
1 -1 V
1 -1 V
2 -1 V
1 -2 V
1 -1 V
2 -1 V
1 -1 V
1 -2 V
2 -1 V
1 -1 V
1 -1 V
2 -2 V
1 -1 V
1 -1 V
2 -1 V
1 -1 V
1 -2 V
1 -1 V
2 -1 V
1 -1 V
1 -2 V
2 -1 V
1 -1 V
1 -2 V
1.000 UL
LT2
2007 1099 M
263 0 V
400 990 M
20 0 V
20 -1 V
20 0 V
20 -1 V
19 0 V
20 -1 V
20 -1 V
20 0 V
20 -1 V
20 -1 V
20 -1 V
20 -1 V
20 -1 V
20 -1 V
19 -1 V
20 -1 V
20 -2 V
20 -1 V
20 -1 V
20 -2 V
20 -2 V
20 -2 V
20 -2 V
20 -2 V
19 -2 V
20 -3 V
20 -2 V
20 -3 V
20 -3 V
20 -3 V
20 -3 V
20 -4 V
20 -3 V
20 -4 V
19 -5 V
20 -4 V
20 -5 V
20 -5 V
20 -5 V
20 -6 V
20 -6 V
20 -6 V
20 -7 V
20 -7 V
19 -8 V
20 -8 V
20 -8 V
20 -9 V
20 -9 V
20 -10 V
20 -10 V
20 -11 V
20 -11 V
20 -11 V
19 -13 V
20 -12 V
20 -14 V
20 -14 V
20 -14 V
20 -15 V
20 -16 V
20 -16 V
20 -17 V
20 -17 V
19 -18 V
20 -19 V
20 -19 V
20 -20 V
20 -20 V
20 -21 V
20 -22 V
20 -22 V
20 -23 V
20 -23 V
19 -24 V
20 -25 V
20 -25 V
20 -25 V
20 -26 V
20 -27 V
20 -27 V
20 -28 V
15 -21 V
1.000 UL
LT3
2007 999 M
263 0 V
400 1101 M
20 -3 V
20 -3 V
20 -4 V
20 -3 V
19 -3 V
20 -4 V
20 -3 V
20 -4 V
20 -4 V
20 -5 V
20 -4 V
20 -5 V
20 -5 V
20 -5 V
19 -6 V
20 -6 V
20 -6 V
20 -7 V
20 -7 V
20 -8 V
20 -8 V
20 -9 V
20 -10 V
20 -11 V
19 -12 V
20 -13 V
20 -15 V
20 -17 V
20 -19 V
20 -23 V
20 -26 V
20 -33 V
20 -43 V
20 -58 V
19 -94 V
20 -230 V
stroke
grestore
end
showpage
}}%
\put(1957,999){\makebox(0,0)[r]{Asymptote (9)}}%
\put(1957,1099){\makebox(0,0)[r]{Asymptote (10)}}%
\put(1957,1199){\makebox(0,0)[r]{HYG}}%
\put(1957,1299){\makebox(0,0)[r]{Exact}}%
\put(1385,350){\makebox(0,0){$L/R-1$}}%
\put(200,806){%
\makebox(0,0)[b]{\shortstack{$A_0(L/R)-A_0(\infty)$}}%
}%
\put(2370,100){\makebox(0,0){ 10}}%
\put(1713,100){\makebox(0,0){ 1}}%
\put(1057,100){\makebox(0,0){ 0.1}}%
\put(400,100){\makebox(0,0){ 0.01}}%
\put(350,1412){\makebox(0,0)[r]{ 10}}%
\put(350,1008){\makebox(0,0)[r]{ 1}}%
\put(350,604){\makebox(0,0)[r]{ 0.1}}%
\put(350,200){\makebox(0,0)[r]{ 0.01}}%
\end{picture}%
\endgroup
 
} 
{\small Fig.~1. $A_0(L/R)-A_0(\infty)$
  as a function of the relative separation $L/R$ calculated by
  different methods.}\\

Although one can not expect that the set of ${F_n,s_n}$ given by HYG
(\ref{F_n}),(\ref{s_n_L}) would produce, upon substitution into
(\ref{Ck_def}), the correct expansion coefficients $A_k$ for $k>0$, it
is still possible that the value of $A_0$ obtained in this manner is
correct. However, as is demonstrated in Fig.~1, this is not so.  In
this figure, we plot the function $A_0(L/R)-A_0(\infty)$ (for
conducting spheres, $A_0(\infty)=3$) calculated by different methods.
The mathematically rigorous result is shown by the solid curve, and
the result of HYG by the long dash. We also show in this figure two
analytical asymptotes valid for $L>> R$ (shorter dash) and $L
\rightarrow R$ (dots). The different curves in Fig.~1 are explained
below in more detail. At this point, we note that the result of HYG
for $A_0(L/R)$ is accurate at large separations ($L \gg R$), but
breaks down when $L/R \approx 1.2$, and becomes grossly inaccurate at
$L/R\approx 1.03$. In particular, the HYG curve has a singularity at
$L/R=x_c \equiv (2^{2/3}+1)/2^{4/3}\approx 1.026$. This is due to the
fact that the first depolarization factor $s_1$ defined by
(\ref{s_n_L}) crosses zero when $L/R=x_c$. The appearance of negative
depolarization factors for smaller inter-sphere separations is
unphysical and can, in particular, result in divergence of the
electrostatic polarizability~\cite{markel_04_3}.

In the next two paragraphs I explain how the data for different curves
shown in Fig.~1 were calculated. The solid curve was obtained by
diagonalization of the electromagnetic interaction operator $W$ whose
matrix elements are given~\cite{mackowski_95_1} by

\begin{eqnarray}
&& W_{il,i^{\prime}l^{\prime}} = {{l\delta_{l l^{\prime}}\delta_{i
        i^{\prime}}} \over {2l+1}} + \nonumber\\
&& \hspace{1cm} (1-\delta_{i i^{\prime}})(-1)^{l^{\prime}}[{\rm
  sgn}(z_{i}-z_{i^{\prime}})]^{l+l^{\prime}} \nonumber \\
&& \times \sqrt{{l l^{\prime}} \over
  {(2l+1)(2l^{\prime}+1)}} {{(l+l^{\prime})!} \over
  {(L/R)^{l+l^{\prime}+1} l! l^{\prime}!}} \ ,
\label{W_def}
\end{eqnarray}

\noindent
where $i,i^{\prime}=1,2$ label the spheres, $l,l^{\prime}=1,2,\ldots$
and $z_i$ is the $z$ coordinate of the center of $i$th sphere,
assuming the $z$-axis coincides with the axis of symmetry. The
depolarization factors $s_n$ are the eigenvalues of $W$ while the
oscillator strengths can be found as squared projections of the
corresponding eigenvectors $\vert n \rangle$ on the vector of external
field: $F_n =\langle E\vert n\rangle \langle n \vert E\rangle$, where
$\vert E\rangle$ is normalized so that $\langle E\vert E
\rangle=1$~\cite{markel_04_3}. The matrix defined in (\ref{W_def}) was
truncated so that $l,l^{\prime}\leq l_{\rm max}=1000$ and diagonalized
numerically. In the absence of round-off errors and in the limit
$l_{\rm max} \rightarrow \infty$, such diagonalization would produce
the infinite set of exact values $s_n, \ F_n$. We note that at $l_{\rm
  max}=1000$ and $L/R\geq 1.01$, all the modes whose oscillator
strength are not very small (i.e., greater than 0.001) have converged
with a very high precision, and that the round-off errors do not
influence the results in any noticeable way since the matrix $W$ is
well-conditioned.

The dots and short dash in Fig.~1 show the theoretical asymptotes
obtained by Mazets who has derived an expression for $A_0$ in terms of
hypergeometrical functions~\cite{mazets_00_1}. He has also provided
simple asymptotic expansions which are valid for small and large
inter-sphere separations. Thus, for longitudinal excitations,

\begin{eqnarray}
&& A_0 \approx 3\left[ 2\zeta(3) - {{\zeta^2(2)} \over {C +
      \ln\left(2/\sqrt{(L/R)^2-1}\right)}} \right] \ , \nonumber \\
&&\hspace{4cm} L \rightarrow R \ ,
\label{C0_close}
\\
&& A_0 \approx 3\left[1 + {1\over 4}\left({R\over L}\right)^3 +
  {1\over 16} \left({ R \over L} \right)^6 \right] \ ,
\nonumber \\
&&\hspace{4cm} L\gg R \ ,
\label{C0_far}
\end{eqnarray}

\noindent
where $\zeta(x)$ is the Riemann zeta-function and $C$ is the Euler
constant. The second term in the right-hand side of (\ref{C0_far}) is
a correction due to the dipole-dipole interaction while the third term
describe the next non-vanishing input due to the higher multipole
interaction.  It can be verified that the asymptotic expansion of the
HYG result coincides with (\ref{C0_far}) at least up to the sixth
order in $L/R$. However, the small-separation asymptote
(\ref{C0_close}) is dramatically different from the one that follows
from the HYG formulas.

\centerline{\psfig{file=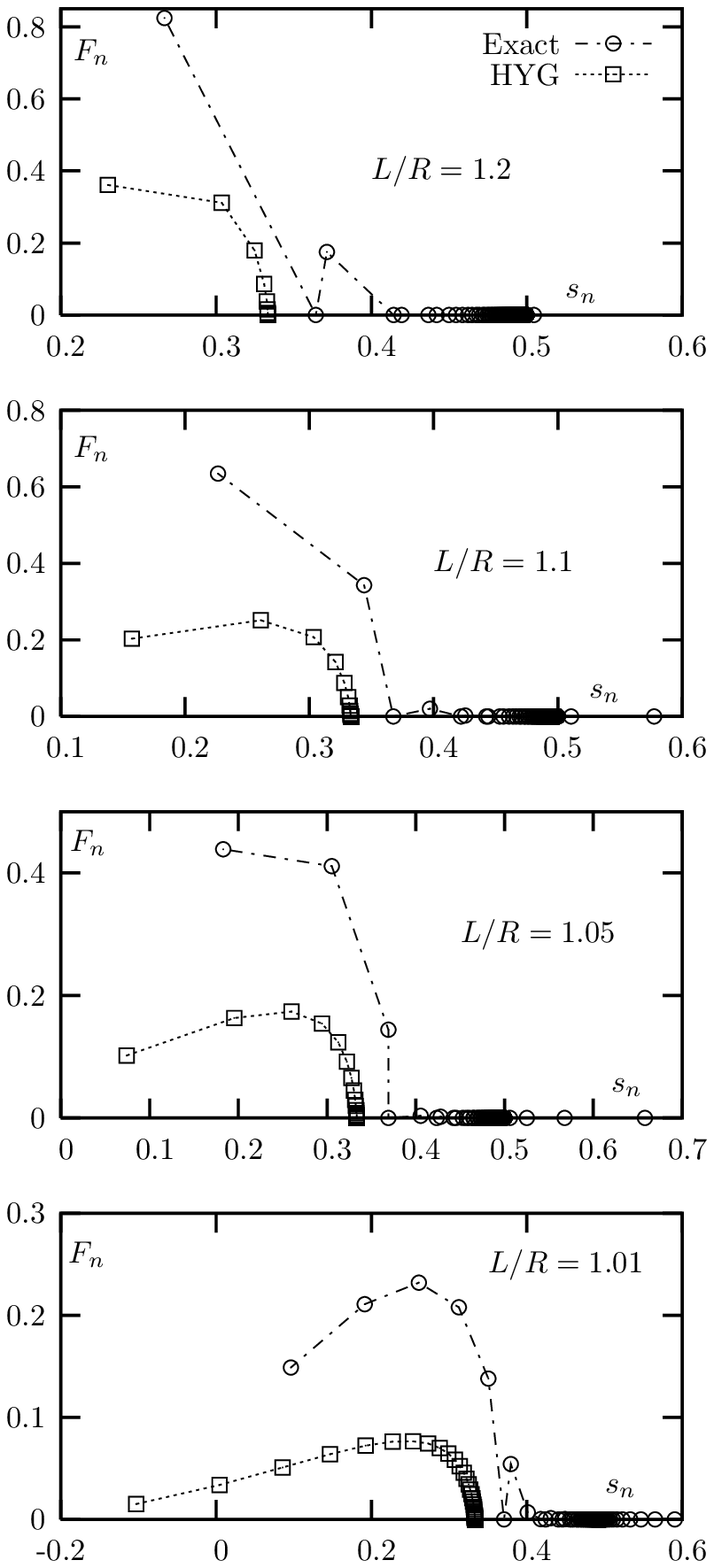,width=8.0cm,bbllx=190bp,bblly=85bp,bburx=426bp,bbury=600bp,clip=t}}
{\small Fig.2 Bergman-Milton depolarization factors, $s_n$, and the
  corresponding oscillator strengths, $F_n$, for different relative
  inter-sphere separations $L/R$. Dashed lines are plotted to guide
  the eye.}\\

Next, we compare the coefficients $s_n,F_n$ defined by
(\ref{F_n}),(\ref{s_n_L}) according to HYG with respective values
obtained by direct diagonalization of the interaction matrix $W$. The
results are shown in Fig.~2. A significant discrepancy already exists
at $L/R=1.2$ and becomes more dramatic as this ratio approaches unity.
Negative depolarization factors are present in the plot for
$L/R=1.01$. We note that the smallest inter-sphere separation
considered by HYG \ was $L/R=1+1/30\approx 1.033$. \ As was mentioned

\centerline{\psfig{file=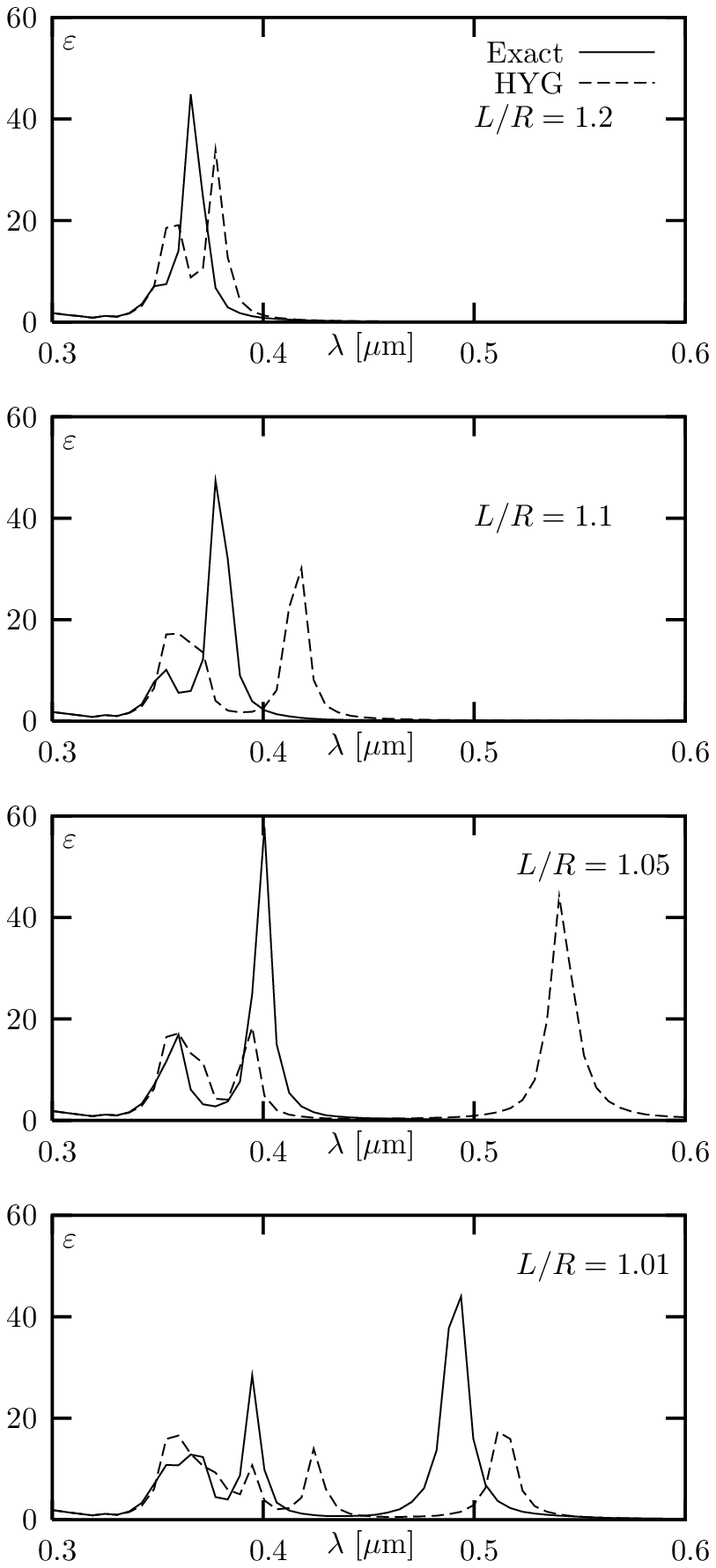,width=8.0cm,bbllx=190bp,bblly=85bp,bburx=426bp,bbury=600bp,clip=t}}
{\small Fig.~3. Dimensionless extinction parameter
  $\varepsilon=\sigma_e/kv$ as a function of wavelength $\lambda$,
  where $\sigma_e$ is the extinction cross section, $k=2\pi/\lambda$,
  $v$ is the total volume of the scatterer, plotted for different
  relative inter-sphere separations $L/R$. Polarization of the
  incident field is parallel to
  the axis of symmetry.}\\

\noindent
above, the negative depolarization factors appear for $L/R
\leq x_c \approx 1.026$. At these separations, results of any
calculation based on the HYG formalism are expected to be grossly
inaccurate and unphysical. However, this fact is not explained in
Refs.~\cite{huang_02_1,huang_04_1}. For example, the choice of values
for $L/R$ in Fig.~5 of Ref.~\cite{huang_02_1} appears to be random,
while, in fact, all these values satisfy the critical condition
$L/R>x_c$.

While it is demonstrated in Fig.~2 that the values of $s_n,F_n$
calculated according to HYG are inaccurate, these coefficients are not
directly measurable in an experiment. However, they can be used to
calculate various physically measurable quantities. For example, the
extinction cross section is given by $\sigma_e = 4\pi k v {\rm Im}
\sum_n F_n/(s + s_n)$. In Fig.~3 we plot the extinction spectra of two
silver nanoparticles obtained for the same inter-sphere separations as
in Fig.~2 and for the longitudinal polarization of the external field.
Interpolated data for silver from Ref.~\cite{johnson_72_1} have been
used to calculate the spectral parameter $s$ as a function of
wavelength. It can be seen that the spectra calculated using the
formulas (\ref{F_n}),(\ref{s_n_L}) for $s_n,F_n$ differ dramatically
from those calculated with the use of exact values of these
coefficients.  The discrepancy is evident even at relatively large
separation, $L/R=1.2$. It should be noted that in the case $L/R=1.01$
the HYG spectra exhibit unlimited growth with the wavelength which
starts in the near-IR region (data not shown). This is due to the
appearance of negative depolarization factors and contradicts the
general sum rules for extinction spectra which imply that $\sigma_e$
must decrease faster than $1/\lambda$ in the limit $\lambda\rightarrow
\infty$~\cite{markel_04_3}. Note that the presence of negative
depolarization factors can result in even more severe anomalies of
extinction spectra in dielectrics whose static dielectric permeability
is positive, as well as the value of $s$ in the limit
$\lambda\rightarrow \infty$.

The papers~\cite{huang_02_1,huang_04_1} contain a number of other less
significant inaccuracies. In particular, HYG confuse orientational
averaging (for randomly-oriented bispheres) with the averaging over
polarization of the incident light.  Thus, for example, Eq.~2 in
Ref.~\cite{huang_02_1} is presented as a result of averaging over
polarization for a fixed bisphere.  However, such averaging should
clearly depend on the direction of the incident wave vector relatively
to the axis of symmetry of the bisphere. In fact, the first equality
in this formula gives the result of orientational averaging, except
that HYG are mistaken in stating that $\langle \cos^2\theta \rangle =
\langle \sin^2\theta \rangle = 1/2$.  It is easy to check that
$\langle \cos^2\theta \rangle =1/3$ and $\langle \sin^2\theta \rangle
= 2/3$.  Note that the second equality in Eq.~2 of
Ref.~\cite{huang_02_1} would be correct if the averaging is done over
polarizations of the incident beam for a fixed bisphere, assuming that
the incident wave vector is perpendicular to the axis of symmetry.

It should be noted that on page 4 of Ref.~\cite{huang_02_1}, the
authors acknowledge that the method of images is only approximate but
state that the approximation is very good and make a reference to the
earlier work~\cite{yu_00_1} to support that statement. However, in
Ref.~\cite{yu_00_1} verification of the 
accuracy of the method of images is only done for relatively large
separations, namely $L/R\geq 1.5$ (Figs. 3, 4 in Ref.~\cite{yu_00_1}).
At these separations, the multipole effects are generally not
important, which clearly follows from the data shown in these figures.
However, in later publications, HYG have used the method for much
smaller separations, typically, $L/R=1+1/30$.

Finally, on the same page of Ref.~\cite{huang_02_1}, the authors
write: ``More accurate calculations based on bispherical coordinates
can be attempted.'' This was, in fact, done in the above-referenced
paper by Mazets~\cite{mazets_00_1}, although only for perfect
conductors.  More general analytical results can be obtained with the
use of the theory of hypercomplex variables (a generalization of the
conformal mapping)~\cite{vagov_94_1}.

\bibliography{abbrevplain,article,local}

\begin{thebibliography}{12}
\bibitem[{fn2()}]{fn2}
  \bibinfo{note}{A slight change of notations compared to those of HYG
  is adopted in this Comment. Namely, the papameter $\alpha$ of HYG is denoted
  by $a$ in order to avoid confusion with the polarizability; the spectral
  parameter is defined as $s=1/(\epsilon-1)$ instead of $s=1/(1-\epsilon)$ and
  the coefficients $F_n$ (\ref{F_n}) differ from those of HYG by the overall
  factor $-3$.}
\end{thebibliography}

\end{document}